\documentclass[runningheads]{llncs}
\usepackage{graphicx}
\usepackage[colorlinks=true,linkcolor=blue]{hyperref}
\usepackage{subcaption}

\usepackage{amsfonts}
\usepackage{amsmath}
\DeclareMathOperator*{\argmin}{arg\,min}
\usepackage{array}
\usepackage{multirow}

\begin{document}
\title{Implicit U-Net for volumetric medical image segmentation}
\author{Sergio {Naval Marimont}\inst{1} \and
Giacomo Tarroni\inst{1, 2}}
%

% index{Naval Marimont, Sergio} 
% index{Tarroni, Giacomo} 

\authorrunning{S. Naval Marimont and G. Tarroni}
\institute{CitAI Research Centre, City, University of London, London, UK \and
BioMedIA, Imperial College, London, UK
\email{\{sergio.naval-marimont,giacomo.tarroni\}@city.ac.uk}}
\maketitle              % typeset the header of the contribution
\begin{abstract}

U-Net has been the go-to architecture for medical image segmentation tasks, however computational challenges arise when extending the U-Net architecture to 3D images. We propose the Implicit U-Net architecture that adapts the efficient Implicit Representation paradigm to supervised image segmentation tasks. By combining a convolutional feature extractor with an implicit localization network, our implicit U-Net has 40\% less parameters than the equivalent U-Net. Moreover, we propose training and inference procedures to capitalize sparse predictions. When comparing to an equivalent fully convolutional U-Net, Implicit U-Net reduces by approximately 30\% inference and training time as well as training memory footprint while achieving comparable results in our experiments with two different abdominal CT scan datasets.

\end{abstract}
\section{Introduction}
U-Net \cite{Ronneberger2015} is the go-to architecture for medical image segmentation tasks \cite{Litjens2017}. A U-Net consists of two convolutional networks an encoder or feature extraction  network and a decoder or localization network. U-Net incorporates \textit{skip connections} that share feature maps directly from encoder to decoder layers with the same spatial resolution. 

Different approaches have been proposed to extend U-Nets to volumetric images. 3D convolutions, as used in V-Net \cite{Milletari2016}, are very computationally challenging at training and inference time \cite{Hesamian2019}. Despite this limitation, U-Nets with fully 3D convolutional architectures have been the building block of general purpose segmentation techniques such as \cite{Jin2020,Isensee2021} that have been state-of-the art until recently. Lately substantial changes to the network design have been made: e.g. Cotr \cite{Xie2021cotr} and UNETR \cite{hatamizadeh2022unetr} improve U-Net performance by replacing some of the convolutions with Transformers in the architecture. 

In a related research direction, there is a growing interest in reducing computational requirements towards improved practical application of deep learning in medical image analysis. Lighter and faster models lead to faster experiments to test achievable segmentation performance, faster hyper-parameter tuning; faster inference time is highly beneficial when processing large medical datasets \cite{sudlow2015uk}; smaller memory footprint translates to better model portability and lower hardware requirements for hospitals and companies running actual segmentation tasks. \cite{Zhang2021} won the MICCAI FLARE 2021 \footnote{\href{https://flare.grand-challenge.org/FLARE21/}{\texttt{Fast and Low GPU memory Abdominal oRgan sEgmentation https://flare.grand-challenge.org/FLARE21/}}} challenge with a U-Net architecture in which 3D decoder convolutions were replaced by anisotropic convolutions and a two step, coarse-to-fine segmentation framework. A different strategy to improve efficiency is to integrate features extracted with 2D convolutions from three orthogonal views of a volumetric image. This strategy is generally referred to as 2.5D convolution and it has been widely explored in segmentation methods \cite{Prasoon2013,Moeskops2016,Roth2018}. However, 2.5D approaches are lagging behind in performance and they do not leverage the availability of volumetric data \cite{Hesamian2019}. \cite{Dong2020} proposes to improve efficiency by training a 3D encoder network to predict the transformations required to compare the volume against an atlas and thus associate a segmentation mask. This approach is largely efficient given that it removes most of the computation associated with the 3D decoder branch, however it requires additional validation in image modalities with high inter-subject variance such as abdominal CT scans.  \cite{Alalwan2021} proposed to use depth-wise separable convolutions in the encoder network to improve the overall efficiency.

\subsubsection{Implicit representation:} 
Implicit Field learning (or occupancy networks) has been recently proposed for 3D shape representation. Implicit networks can reconstruct 3D shapes using their implicit surface representation \cite{mescheder2019occupancy,chen2019learning}. Instead of using convolutional architectures to generate dense voxel outputs, a linear neural network learns to classify as background or object the spatial coordinates \cite{mescheder2019occupancy,chen2019learning}. In a related approach, \cite{park2019deepsdf} learns the signed distance function with respect to the object surface instead of performing object/background classification. 

Implicit approaches can be more efficient than voxel representations in 3D images because they can generate sparse outputs (i.e. for only a subset of points), while in voxel representations, number of parameters and computation grow with a cubic function of the image size. Some recent research has proposed to leverage Implicit Fields in medical image analysis, including unsupervised anomaly detection \cite{naval2021implicit} and super resolution \cite{irem}. 

\subsubsection{Contributions:} 
The Implicit Field formalism has allowed to avoid the use of convolutional layers and still produce smooth(er) outputs for shape reconstruction. Following the previously described research direction of making changes to the UNet architecture, we explored what could be achieved by removing the Decoding branch of the UNet, aiming for a more lightweight and faster to train design, without sacrificing accuracy.

We propose a new architecture, named Implicit U-Net, that adapts the principles of the efficient implicit representation paradigm to supervised volumetric medical image segmentation. Given that the Implicit decoder enables prediction of only a subset of points, we evaluate mechanisms to leverage this feature to improve the efficiency of both training and inference. Our main contributions are the following:
\begin{itemize}
    \item We propose an adaptation of the implicit architecture for segmentation tasks whereby the Implicit decoder receives features from multiple spatial resolutions, replicating the function of skip-connections in a standard U-Net;
    \item We introduce training and inference procedures that leverage sparse predictions as a strategy to improve training and inference efficiency, obtaining important reductions in time and memory footprint;
    \item We tested this approach in two datasets from the Medical Segmentation Decathlon \cite{antonelli2021medical} and achieved accuracy comparable to the standard U-Net while reducing the inference and training time as well as memory footprint by approximately 30\%.
\end{itemize}

\section{Methods}
\subsubsection{Implicit field representation:} 3D images are commonly represented as a dense set of voxels. Implicit field networks represent images by learning a continuous mapping $f$ between spatial coordinates $\mathbf{p}=(x,y,z) \in \mathbb{R}^3$ and a target variable $T$. Therefore, implicit networks receive as inputs and make predictions for a sparse set of points. We leverage this feature to reduce computation during training and inference time.

In addition to spatial coordinates, features $\mathbf{z} \in \mathbb{R}^D$  describing the image are also received as inputs by the Implicit decoder network:

\begin{equation} \label{eq:1}
    f: \mathbb{R}^3 \times \mathbb{R}^D \to T
\end{equation} 

In a segmentation task $f$ learns the posterior probability over $C$ objective classes for continuous spatial coordinates $\mathbf{p}$ and latent features $\mathbf{z}$. In binary segmentation, we model the posterior probability using the logistic sigmoid activation function. Training an implicit network generally implies minimizing the training loss $\mathcal{L}$ over a set of $k$ points which are randomly sampled from $N$ images. For an implicit network parametrized by $\theta$, \ref{eq:2} describes the training objective:

\begin{equation} \label{eq:2}
    \argmin_{\theta}
    \sum^N_{i=1}{\left(\sum^K_{j=1}{\mathcal{L}\left(f_\theta\left(\mathbf{z}_i,\mathbf{p}_{i,j}\right),t_{i,j}\right)}\right)}
\end{equation}

In our implicit decoder we used the architecture proposed in \cite{park2019deepsdf}. The decoder is a feed-forward network composed of 8 fully-connected layers with all hidden layers with 512 units, ReLU as activation, weight normalization and dropout 0.2 in all layers. Prior to feeding coordinates $(x,y,z)$ to the decoder, the coordinates are normalized to the range $[-1,1]$  and then encoded using function described in \cite{mildenhall2020nerf} (we used $L=10$ in our experiments).

\subsubsection{Implicit U-Net architecture:} We propose to use a standard 3D convolutional neural network (CNN) to generate features $\mathbf{z}$ that are typically passed to the decoder. In a standard CNN encoder, spatial dimensions of feature maps are contracted progressively. In our implementation we use 2-strided convolutions for pooling. Consequently, given an image with dimensions $(W, H, D)$, the feature map dimensions in the downward block (identified with index $b=\{0,1,...n\}$) are $(W // 2^b, H // 2^b, D // 2^b)$, $//$ being the floor division operation. 

\begin{figure}[b!]
\includegraphics[width=\textwidth]{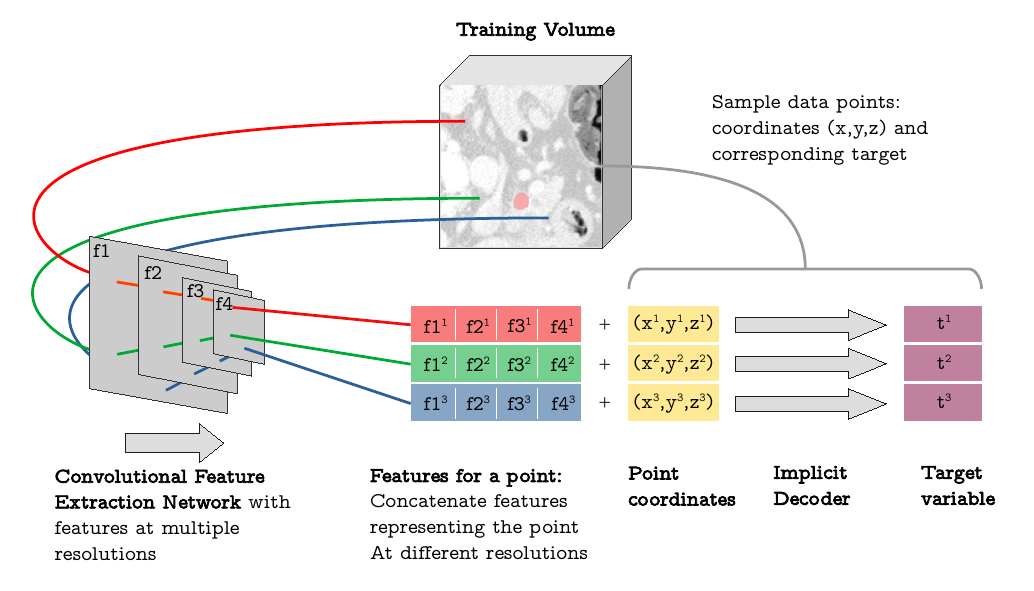}
\caption{Diagram of the gather feature vector operation. For example, for \textbf{p} with coordinates (8,16), in the 2nd feature map get (4,8), in 3rd feature map get (2,4), etc. Features extracted at different resolutions are concatenated with the spatial coordinates and passed to the Implicit decoder.} \label{fig1}
\end{figure}

In the original implicit network implementations, it is proposed to obtain features only from the deepest CNN encoder layer, which is expected to contain global features. However, for segmentation tasks we hypothesized that both global and local features are required to make voxel-wise predictions. Consequently, we propose an architecture that extracts features from multiple spatial resolutions at the same time. In order to achieve this, we concatenate features at each resolution in the encoder network. Specifically, for a point $\mathbf{p}$ with spatial coordinates $(x,y,z)$, we concatenate the vectors with coordinates $(x//2^b,y//2^b,z//2^b)$ from the $b$ feature maps in the CNN encoder. Intuitively, we gather the feature vector in the same relative position as the point in the original image (see Fig. \ref{fig1}). With this approach we intend to not only give the decoder access to local features but to provide signal directly at multiple depths in the CNN, similarly to how \textit{deep supervision} \cite{lee2015deeply} in standard U-Nets operate. 

\begin{figure}
\includegraphics[width=\textwidth]{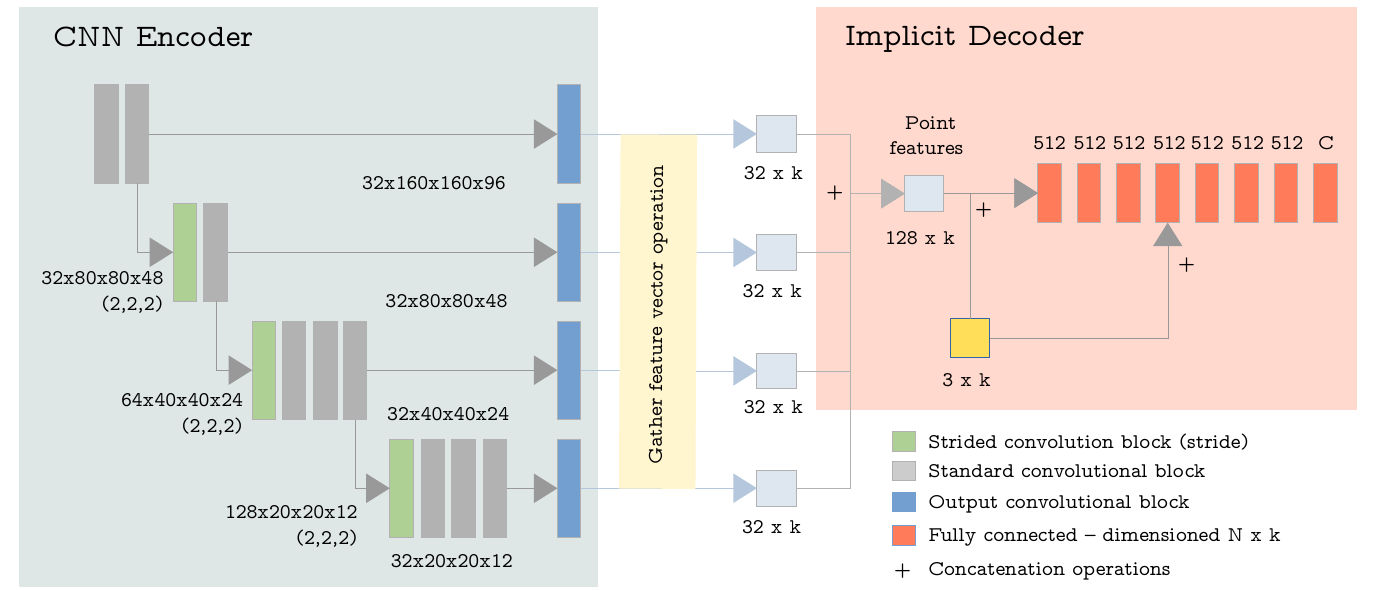}
\caption{Implicit U-Net architecture with CNN encoder and Implicit decoder blocks. Gather feature vector operation selects the feature vectors at the relative position of the point with respect the original image.} \label{fig2}
\end{figure}

We adapted the encoder architecture from \cite{Isensee2021}, using in all our experiments 2 blocks of 2 convolutions followed by 2 blocks of 4 convolutions. We also evaluated the original architecture of 5 blocks of 2 convolutions but found that it produced \textit{patchy} artefacts.  We hypothesised that the sharp edges in the artefacts are related to the gather feature vectors operation and that the sharp straight edges are produced in the image areas where there is a transition from one deep latent vector to the next. We therefore use a fewer number of blocks but with more convolutions per block to effectively increase the receptive field of the 4th block. This subtle change in the encoder architecture reduced the artefacts because each of the deepest feature vectors is used in the predictions of smaller areas (16x16x16 voxels) compared to the architecture with 5 blocks (32x32x32 voxels) (Fig. \ref{fig2}).

\subsubsection{Training procedure:} The CNN encoder and Implicit decoder networks are trained simultaneously end-to-end. The training consist of 1) CNN encoder forward pass, 2) random sampling of $k$ points from the image, 3) gather feature vector operation for the sampled points, 4) Implicit decoder forward pass with feature vectors and point coordinates and 5) backward pass. 

\begin{figure}[b!]
\includegraphics[width=\textwidth]{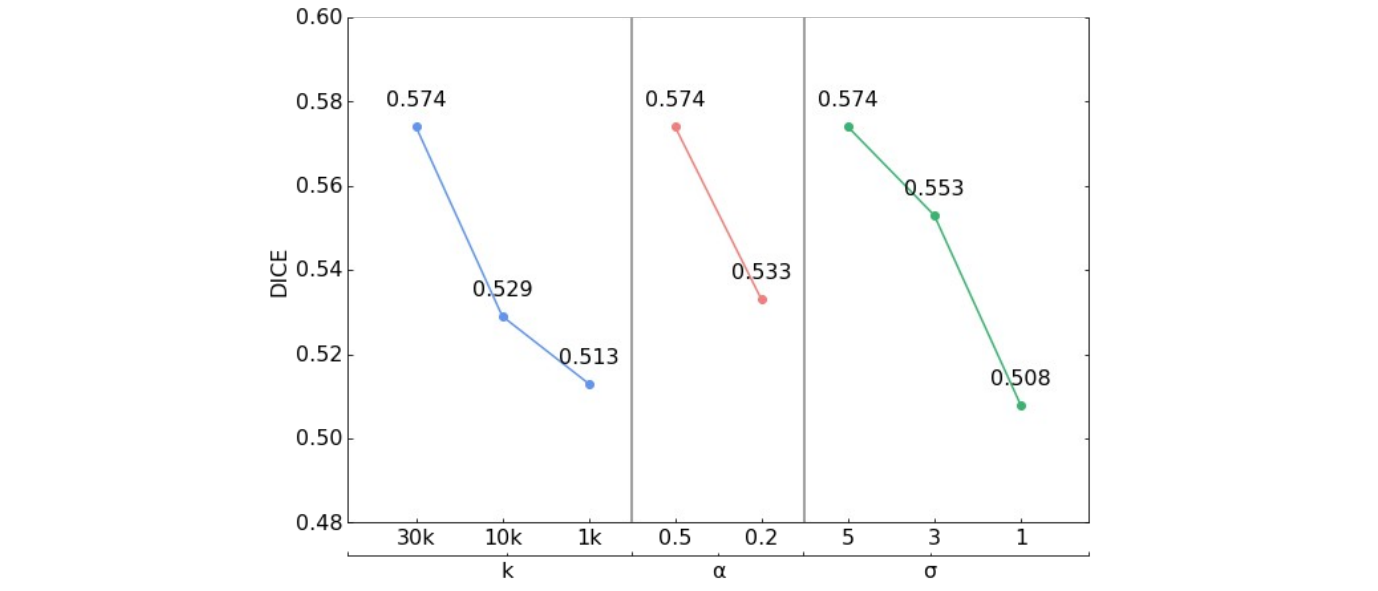}
\caption{Mean DICE score in Lungs dataset for different sets of hyper-parameters, starting from the baseline $k = 30,000$, $\alpha = 0.5$ and $\sigma$ = 5. Results are from preliminary experiments with a batch-size 3 and 1,200 epochs.} \label{fig3}
\end{figure}

In reference to step 2, we hypothesized that model training and final accuracy could be improved by oversampling points near label boundaries. We implemented this strategy with two hyper-parameters. $\alpha = [0,1]$ specifies the proportion of points that are sampled from the label boundary, with $(1-\alpha)$ being the proportion of points being sampled uniformly from the full image.  Secondly, $\sigma$ is used to control the distance from the boundary of the $k \times \alpha$ sampled points. The final points are obtained by adding displacement sampled from $N(0,\sigma)$ to the $k \times \alpha$ points sampled from the boundary. We performed hyper-parameter tuning for $k$, $\alpha$ and $\sigma$ (see results on Fig. \ref{fig3} for the Lungs dataset). The final experiments were run with $k=30,000$, $\alpha = 0.5$ and $\sigma = 5$.

In our experiments we used a patch-based training pipeline, with the spatial coordinate system defined for each patch as opposed to the full image. We sampled patches using 1 to 2 positive to negative ratio. Patch-sizes are described in the Experiments section.

\subsubsection{Inference procedure:} With the objective of improving inference efficiency, we capitalize the feature of Implicit Fields that allows to make predictions for sparse points and propose a three-stage inference process:
\begin{itemize}
    \item \textbf{Broad prediction}: this stage makes predictions for a subset of points in the image forming a broad mesh. In our final implementation the broad mesh was created selecting one in every 4 voxels across all dimensions, thus extracting predictions for $1/4^3=1/64$ of the input voxels. Predictions for the remaining points are obtained through nearest neighbour interpolation;
    \item \textbf{Fine boundary prediction}: it consists of 2 steps: 1) we localise a predicted segmentation boundary using the initial broad prediction; and 2) If a boundary is identified, the points near the broad boundary are predicted by the Implicit decoder to obtain the finer details around the segmentation boundary.
\end{itemize}

The proposed inference procedure adds as hyper-parameter the spacing of voxels taken in the broad prediction. We evaluated 2, 4 and 8 spacings. With a spacing of 4 we did not observe any difference in DICE score compared to predicting every single voxel and was the fastest of the three values evaluated. 

Also inference is patch-based and uses a sliding-window approach with an overlap of 0.3. Patch predictions are consolidated using Gaussian weighting. Predictions are finally smoothed using a 3D average pooling filter with a kernel of size 3.

\section{Experiments and Results}
\subsection{Experimental set-up:} We evaluated our Implicit-U-Net on two datasets from the Medical Segmentation Decathlon \cite{antonelli2021medical}, specifically Pancreas and Lung CT scans: 

\begin{itemize}
    \item \textbf{Task06 Lung dataset}: 64 abdominal CT scans. The task is to identify the lung tumor mass. Images were resampled to a common 1.25mm resolution along the the z-axis. 
    \item \textbf{Task07 Pancreas dataset}: 282 contrast-enhanced, abdominal CT scans. The task is to identify pancreas (label 1) and pancreatic tumor mass (label 2). Images were resampled to a common 2.5mm resolution along the z-axis. 
\end{itemize}

Image intensities were capped at the 95th percentile dataset-wise. Subsequently we applied z-score intensity normalization. Training and evaluation were both performed patch-wise with 160x160x96 patches in both abdominal CT datasets. Training augmentations included elastic transforms, scaling, rotations, axis flip, gaussian noise and random contrast. 

We compared our model with an equivalent standard 3D U-Net architecture as implemented in \cite{Isensee2021}. Both standard U-Net and ours were trained using a combined DICE and Cross-Entropy loss, and AdamW optimizer with a learning rate of $10^{-4}$ and a batch-size of 6 for a fixed number of epochs (2,000 in Lungs and 800 in Pancreas). 

5-fold cross validation was used in the two datasets. The reported results show  the mean validation set performance for the best performing models fold-wise.

Network implementation, training and testing procedures are made publicly available in \footnote{\href{https://github.com/snavalm/imunet_miua22}{\texttt{https://github.com/snavalm/imunet\_miua22}}}. Experiments were run across multiple hardware, including Nvidia 2 x RTX2080Ti, TITAN RTX and A100.

\begin{table}
\caption{Quantitative results on Segmentation Decathlon datasets.}\label{Table:tab1}
\subcaption*{\textbf{Lung}}
\begin{tabular}{p{1.5cm} >{\centering\arraybackslash}p{5.6cm} >{\centering\arraybackslash}p{1.5cm} >{\centering\arraybackslash}p{1.5cm}  >{\centering\arraybackslash}p{1.5cm}}
% \begin{tabular}{ m{3.5cm} m{2.3cm} m{2.1cm} m{1.0cm} m{1.5m} c }
% \toprule[1.5pt]
\hline
Method &  DICE & Inf t$^1$ & Tr t$^2$ & Tr Mem$^2$ \\
\hline
UNet $^3$ & 64.1 $\pm$ 27.3 & 36.7 & 58.2 & 16,060 \\
im-UNet & \textbf{65.7} $\pm$ 23.3 & \textbf{25.2} & \textbf{41.7} & \textbf{11,728} \\
\hline
\end{tabular}
\subcaption*{\textbf{Pancreas}}
\begin{tabular}{p{1.5cm} >{\centering\arraybackslash}p{2.7cm} >{\centering\arraybackslash}p{2.8cm} >{\centering\arraybackslash}p{1.5cm} >{\centering\arraybackslash}p{1.5cm}  >{\centering\arraybackslash}p{1.5cm}}
% \begin{tabular}{ m{2.5cm} m{2.1cm} m{2.1cm} m{1.0cm} m{1.5m} c }
% \toprule[1.5pt]
\hline
Method &  DICE 1 & DICE 2 & Inf t$^1$ & Tr t$^2$ & Tr Mem$^2$ \\
\hline
UNet $^{3}$ & \textbf{75.8} $\pm$ 9.1 & \textbf{35.5} $\pm$ 29.6 & 9.2 & 59.7 & 16,151 \\
im-UNet & \textbf{75.8} $\pm$ 9.1 & 33.3 $\pm$ 29.2 & \textbf{6.7} & \textbf{44.4} & \textbf{11,459} \\
\hline
\multicolumn{6}{l}{1 - Mean inference time for N=10 images in seconds using sliding window.} \\
\multicolumn{6}{l}{2 - GPU time in seconds and peak memory usage for 100 training steps on 2xRTX2080Ti} \\
\multicolumn{6}{l}{using mixed-precision.} \\
\multicolumn{6}{l}{3 - 3D UNet architecture with deep-supervision as in \cite{Isensee2021}.} \\
\end{tabular}
\end{table}

\subsection{Results and Discussion:}
Table \ref{Table:tab1} contains quantitative results for the experiments in the two datasets. Mean DICE score for each label and standard deviation across subjects are reported as performance metrics. Additionally, mean inference time, training time and peak GPU memory usage for 100 training steps are reported.

\begin{figure}[t!]
\includegraphics[width=\textwidth]{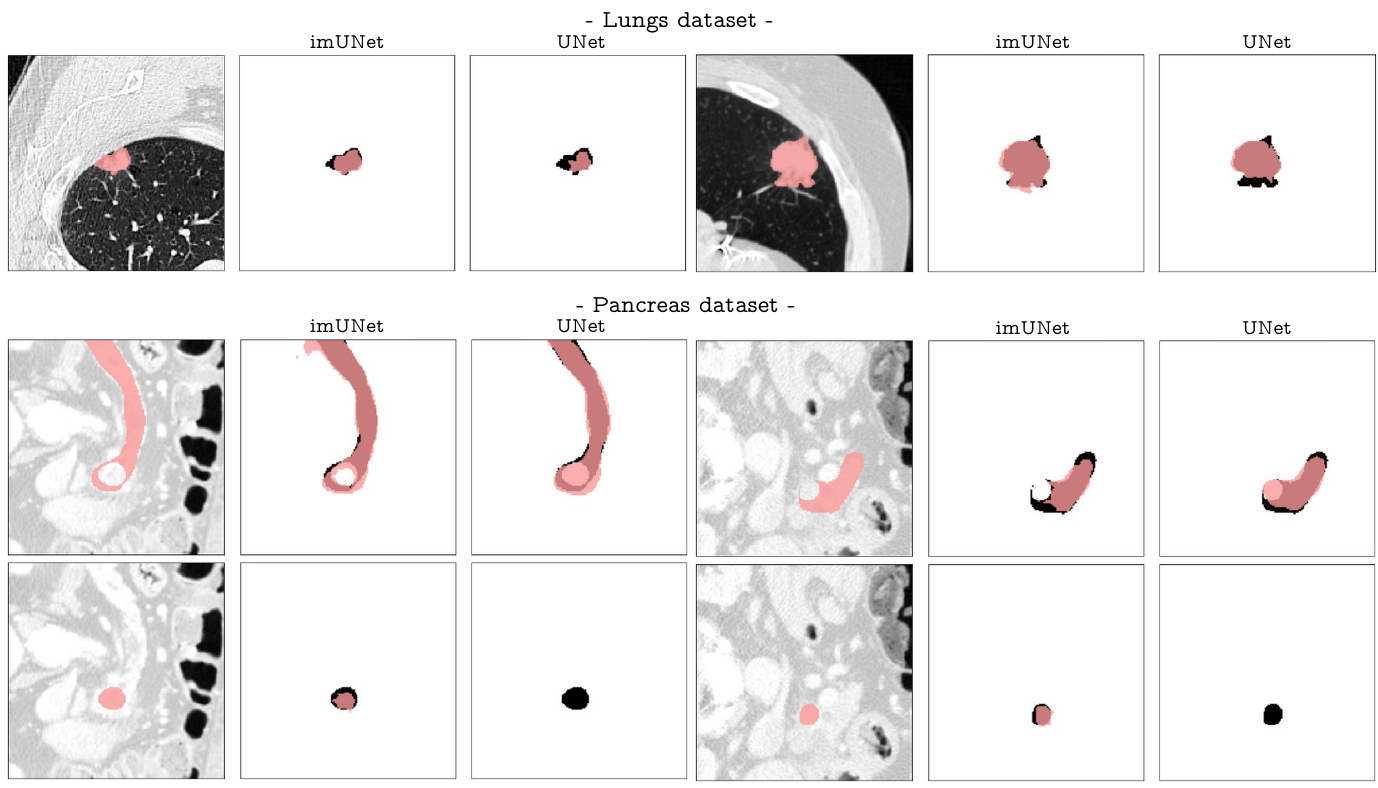}
\caption{Qualitative comparison of Implicit U-Net and standard U-Net in 2 different lungs images and pancreas image. For each image, columns show image and ground truth, Implicit U-Net prediction vs ground truth and U-Net prediction and ground truth. In Pancreas, first row correspond to the label 0 (Pancreas) and second row to label 1 (Tumor).} \label{fig4}
\end{figure}

Fig. \ref{fig4} shows qualitative comparisons in both dataset between segmentation outputs obtained with the the proposed Implicit U-Net (im-UNet) and the standard U-Net.

Implicit U-Net achieves comparable performance to U-Net in both datasets. Our technique seems to outperform the U-Net in the Lungs dataset while underperforming in tumor segmentation in the Pancreas dataset, however in all datasets performance differences are well below one standard deviation. Of note, the relatively low DICE scores and high standard deviations (e.g. in the Lungs dataset) reflect the difficulty of performing the required segmentation task. These datasets where in fact selected to provide a good testbed for our comparisons. Differently from DICE scores, inference, training time and training memory show instead a clear advantage of the Implicit U-Net over the standard U-Net, with reductions in the range of 30 - 35\%. It is important to note that inference time in implicit U-Net depends on the characteristics of the lesion because of the number of fine-boundary prediction steps required at inference. 

With regards to the hyper-parameters added by the model, $k$, $\alpha$ and $\sigma$ for training and the broad-mesh scale for inference, we found that the final parameters proposed appeared optimal in the two Abdominal CT scan datasets evaluated. Further evaluation would be required in very different image modalities or targets.

\section{Conclusion}

We introduced a new strategy to improve efficiency of deep learning architectures in 3D segmentation tasks, which consists in leveraging sparse predictions in an Implicit network that replaces the standard convolutional decoder network. Our experiments show that our method achieves competitive results when compared to the reference architecture for this task (i.e. the 3D U-Net) while improving training and inference times by 30\%. Training time and memory advantages lead to faster research iterations and hyper-parameter tuning. Faster inference makes Implicit U-Net relevant in the current practice dealing with larger datasets, larger image sizes and for hospitals and companies running actual segmentation tasks.

We propose Implicit U-Net in the context of growing research interest to improve U-Net architecture replacing some of the computationally expensive 3D convolutions and we believe that is complementary to other methods such as UNETR that focuses on the encoder architecture. 

Future research will focus on further uncovering the power of the implicit field representations and testing the proposed approach on other datasets/settings. We will also exploring combinations of implicit decoders with transformer encoders, potentially speeding up and improving the model footprint of high-parameter architectures like UNETR.
 
%
% the environments 'definition', 'lemma', 'proposition', 'corollary',
% 'remark', and 'example' are defined in the LLNCS documentclass as well.
%

\bibliographystyle{splncs04}
\bibliography{ref.bib}

\end{document}